\documentclass[final,11pt,twocolumn]{article}

 \usepackage{graphicx}
\usepackage{amssymb}
\usepackage{lineno}

\title{Bucking Coil Implementation on PMT \\ for Active Cancelling of Magnetic Field}

\author{
T.~Gogami$^{a}$, 
A.~Asaturyan$^{b}$, 
J.Bono$^{c}$, 
P.Baturin$^{c}$,
C.~Chen$^{d}$, 
A.~Chiba$^{a}$, 
N.~Chiga$^{a}$, 
Y.~Fujii$^{a}$,\\
O.~Hashimoto$^{a}$\footnote{Deceased},
D.~Kawama$^{a}$\footnote{Current Address : Institute of Chemical and Physical Research (RIKEN), Wako, Saitama, 351-0198,Japan},
T.~Maruta$^{a}$\footnote{Current Address : Accelerator Laboratory, High Energy Accelerator Research Organizaion (KEK), Tsukuba, 305-0801, Japan},
V.Maxwell$^{c}$,
A.~Mkrtchyan$^{b}$,
S.~Nagao$^{a}$,\\
S.~N.~Nakamura$^{a}$,
J.~Reinhold$^{c}$,
A.~Shichijo$^{a}$,
L.~Tang$^{d}$,
N.~Taniya$^{a}$,
S.~A.~Wood$^{e}$,
Z.~Ye$^{d}$\\
a)~Graduate School of Science, Tohoku University, Sendai, 980-8578, Japan\\
b)~Yerevan Physics Institute, Armenia\\
c)~Department of Physics, Florida International University, Miami, Florida 33199, USA\\
d)~Department of Physics, Hampton University, Virginia 23668, USA\\
e)~Thomas Jefferson National Accelerator Facility, Newport News, Virginia 23606, USA}

\newcommand{\cerenkov}{\v{C}erenkov}
\date{}
\begin{document}

\maketitle

\begin{abstract}
Aerogel and water {\cerenkov} detectors were employed to tag kaons for
a $\Lambda$ hypernuclear spectroscopic experiment which used
the $(e,e^{\prime}K^{+})$ reaction in experimental Hall C
at Jefferson Lab (JLab E05-115).
%Fringe fields from the Kaon spectrometer magnet yielded $\sim$6 Gauss
Fringe fields from the kaon spectrometer magnet yielded $\sim$~5 Gauss
at the photomultiplier tubes ( PMT ) for these detectors which could not be
easily shielded.  As this field results in a lowered kaon detection efficiency,
we implemented a bucking coil on each photomultiplier tubes to actively
cancel this magnetic field, thus maximizing kaon detection efficiency.
\end{abstract}

%\begin{keyword}
%%% keywords here, in the form: keyword \sep keyword
%Active magnetic field canceling, bucking coil
%%% MSC codes here, in the form: \MSC code \sep code
%%% or \MSC[2008] code \sep code (2000 is the default)
%\end{keyword}

%\end{frontmatter}

%\linenumbers

\section{Introduction}
\label{lab:intro}

We performed a $\Lambda$ hypernuclear spectroscopic measurement, 
which used the $(e,e^{\prime}K^{+})$ reaction at 
JLab Hall-C in 2009 (JLab E05-115~\cite{cite:proposal}\cite{cite:osamu}).
The scattered electron ($e^{\prime}$) and kaon ($K^{+}$) were
measured in coincidence using the HES ( High resolution Electron 
Spectrometer ) 
and HKS ( High resolution Kaon Spectrometer ), dedicated electron and
kaon spectrometers.
%Momentum vectors of  $e^{\prime}$ and  $K^{+}$ 
%at the target are derived form 
%the measurement of position and angle of the particles 
%at reference planes with spectrometers, HES and HKS, respectively. 
%Then, missing mass can be calculated.
The positions and angles of the scattered particles were measured
with tracking detectors in the focal planes of each spectrometer and
this information was used to reconstruct a missing mass.

\begin{figure}[!htbp]
\begin{center}
\includegraphics[width=8.5cm]{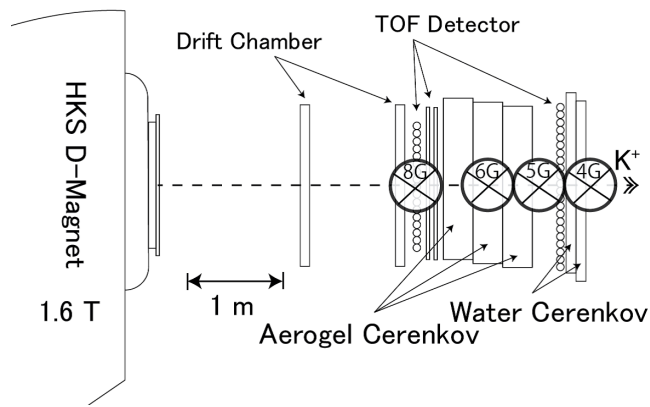}
\caption{A schematic drawing of the HKS detectors of JLab E05-115.
The fringe field of HKS dipole magnet ( normal conducting magnet, 1.6~T ) 
which is 4$\sim$6 G around the {\cerenkov} detectors effects on 
their PMT performance.}
\label{fig:hks_setup}
\end{center}
\end{figure}

However, HKS, the kaon spectrometer, had a very large background of
protons and positive pions.
An aerogel {\cerenkov} detector and a water {\cerenkov} detector, used
to reject $\pi ^{+}$s and protons both on-line and off-line, played
an important role in $K^{+}$ identification.  These {\cerenkov} detectors 
were mounted close to the HKS dipole magnet
(Figure~\ref{fig:hks_setup}) in order to minimize loss of kaons due to decay.
Therefore, fringe field of up to $\sim$~5~G were observed in the
%Therefore, fringe field of up to $\sim 6$ G were observed in the
region of the photomultipler tubes (PMTs) on the {\cerenkov} detectors.
It was expected that the $K^{+}$ identification efficiency would
be reduced because of the effects of magnetic fields on PMT performance.
A previous experiment (JLab
E01-011~\cite{cite:lulin}\cite{cite:tang}\cite{cite:osamu2}\cite{cite:nue}
in 2005) used iron shields around the PMTs.  However, this was not
optimal as it is difficult to shield magnetic fields parallel to the axis
(perpendicular to the face) of the tube and required installation 
of large amount of iron. Therefore, for this
experiment
we implemented a bucking coil system to more effectively cancel out
the fringe fields.

%%%%%%%%%%%%%%%%%%%%%%%%%%%%%%%%%%%%%%%%%%%%%%%%%%%%%%%%%
\section{{\cerenkov} detectors}
\label{lab:Cherenkov}
\subsection{{\cerenkov} detectors in HKS}

\begin{figure}[htbp]
\begin{center}
\includegraphics[width=8cm]{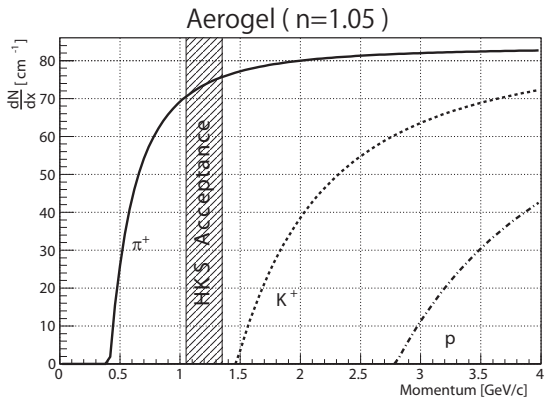}
\caption{Number of {\cerenkov} photons per centimeter vs
%momenta in the radiation medium, aerogel (n=1.055, top figure) and water (n=1.33, bottom figure).}
momentum in the aerogel medium ( n=1.05 ).  Only $\pi^{+}$ generate {\cerenkov}
light within the HKS 
momentum acceptance. The aerogel {\cerenkov} detector is used as a veto
detector to suppress $\pi^{+}$ particles.}
\label{fig:np_aerogel}
\end{center}
\end{figure}

\begin{figure}[!htbp]
\begin{center}
\includegraphics[width=8cm]{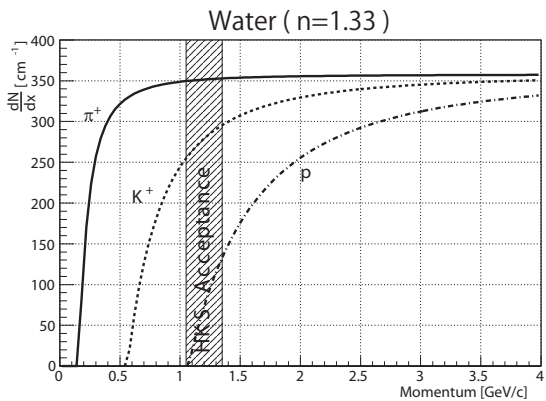}
\caption{Number of {\cerenkov} photons per centimeter vs
%momenta in the radiation medium, aerogel (n=1.055, top figure) and water (n=1.33, bottom figure).}
momentum in water (n=1.33).   
The water {\cerenkov} detector is used to separate $p$ from other particles
by choosing an optimized threshold of number of photoelectrons.}
\label{fig:np_water}
\end{center}
\end{figure}
Over the HKS momentum acceptance of $\sim$1.05~GeV/$c$ 
to $\sim$1.35~GeV/$c$, the number of background particles 
are 2,000:1 for protons and 10,000:1 for
$\pi^{+}$s relative to kaons.
Using aerogel and water, {\cerenkov} detector index of refraction values
of 1.05 and 1.33 were chosen.  
Figures~\ref{fig:np_aerogel} and~\ref{fig:np_water} show predicted {\cerenkov}
light photon yields for the three particle species in the two radiator
media calculated by:
\begin{equation}
\frac{d^{2}N}{dxd\lambda } = \frac{2\pi \alpha z^{2}}{\lambda ^{2}}
\Bigl(1-\frac{1}{\beta ^{2}n^{2}(\lambda)}\Bigr)
\end{equation}
where $N$ is the number of photons, $\alpha$ is the fine structure constant, $\lambda$ 
is the wave length of the {\cerenkov} light, $n(\lambda)$ is the refraction index of the 
medium, and $z$ and $\beta$ are the charge and the velocity factor of the incident particle.
In the figures, $n(\lambda)$ is fixed at 1.05 and 1.33 for aerogel and water,
and the yield is integrated over a $\lambda$ range between 
300~nm and 600~nm, corresponding to the sensitive region of the PMTs.  As
the kaon and proton velocities are below threshold in aerogel,
$\pi^{+}$ and lighter particles can be rejected.  The water {\cerenkov} detector is
sensitive to all three particle species, but can distinguish protons and
kaons by the difference in photon yields.
%The threshold of wather 
%Chrenkov detector was set to loose enogh not to cut $K^{+}$ in the trigger level.

%%%%%%%%%%%%%%%%%%%%%%%%%%%%%%%%%%%%%%%%%%%%%%%%%%%%%%%%%%%%%%%
\subsection{Requirements}
%We required that a rejection ratio of 100/1 for $\pi^{+}$ and $p$ 
We required that a on-line rejection ratio of the order of 1/100 for $\pi^{+}$ and $p$ 
%in order to obtain the desired hypernuclear yield and signal to accidental 
to obtain the desired hypernuclear yield and signal to accidental 
background ratio.
The goal of a high background rejection ratios must be
balanced against the desire to maximize the efficiency of 
$K^+$ detection, particularly in the water {\cerenkov} detectors.  This
efficiency is determined by photo-electron collection efficiency of PMTs, 
a property that is strongly affected by magnetic fields.

A Monte Carlo simulation was done to estimate on-line background rejection
ratios and kaon detection efficiency.  This simulation was normalized
with a cosmic ray tests which found, for the water {\cerenkov} detectors, a mean
signal of 54 photoelectrons per cosmic ray.  According to the simulation, 
%a cut that rejects $98\%$ of protons will accept $K^+$ with a $97\%$
two layers of water {\cerenkov} cuts that rejects $99.96\%$ of protons 
will accept $K^+$ with a $94.5\%$ efficiency.
% Why the \sim in front of these percentages.
However, when the mean number of photoelectrons is reduced by $75\%$ 
( mean signal of 13 photoelectrons ),
%the $K^+$ acceptance rate drops to $87\%$ for the same proton
the $K^+$ acceptance rate drops to $75.7\%$ for the same proton
rejection ratio, directly reducing the yield for $\Lambda$ hypernuclei.
%In this case, beam time of 110$\%$ have to be required to reach same 
%hypernucei yield.  
Given the limits on beam time, efforts to mitigate the effect of
magnetic field on kaon detection efficiency are justified.
% One might say that the graphs presented later always show less than
% a 75% reduction in photoelectron yield, so the loss in hypernuclear
% yield is never as bad as 10%.  

In order to maximize the yield of $\Lambda$ hypernuclei and 
the signal to noise ratio, we set a goal that magnetic fields
should not reduce photoelectron yields by more than $35\%$.

%\section{Test bench for magnetic field effect on PMT}
%%%%%%%%%%%%%%%%%%%%%%%%%%%%%%%%%%%%%%%%%%%%%%%%%%%%%%%%%%%%%
\section{Test of bucking coil}
\label{lab:mageffect}

\begin{center}
\begin{figure}[!htbp]
\includegraphics[width=9cm]{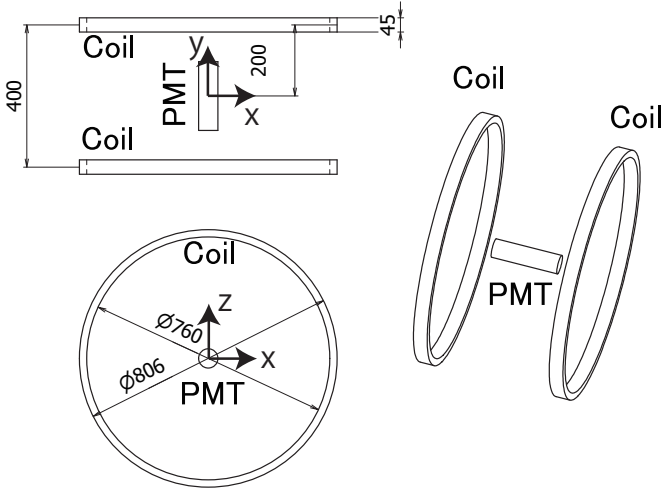}
\caption{A schematic drawing of the setup of bucking coil test. 
An H7195 PMT was placed at the center of the two coils (Helmholtz coil). 
%Dimensions are given in mm.}
Unit in the figure is mm.}
\label{fig:test_setup}
\end{figure}
\end{center}
Before the setup of JLab experiment E05-115 was mounted, 
studies were made of the effects of 
magnetic fields on the {\cerenkov} PMTs and of the ability of bucking 
coils to cancel these effects. 
The test setup is shown in Fig.~\ref{fig:test_setup} which also 
defines the coordinate system. 
A two inches diameter H7195 PMT ( HAMAMATSU, photocathode of bialkali, 
dynode stages of 12, supply voltage of $-2000$~V, typical gain of 3.0$\times$10$^{6}$ ), 
%the tube used in the water {\cerenkov}
which was used in the water {\cerenkov} 
detector, was tested.

%SAW:  I don't think it is necessary to describe the Helmholtz coil so
% much.  Just give the coil radius, coil separation, and note that the
% field is uniform to X% in the region that the tubes are placed.
% B/I plots, comparison to Tosca, plots of field uniformity are not
% needed.

\subsection{Magnetic field generation}
\begin{figure}[htbp]
\begin{center}
\includegraphics[width=8cm]{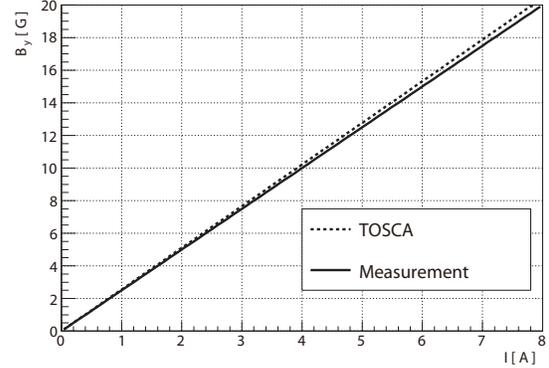}
\caption{Magnetic field ( B$_{y}$ ) and current ( I ) correlation at the 
center of the coils which are measured by 
the hall probe and calculated by OPERA3D-TOSCA\cite{cite:tosca}.}
\label{fig:by_coil}
\end{center}
\end{figure}

A Helmholtz coil which can generate magnetic 
fields to $\sim $16 G was constructed.
The cross section and the number of turns of 
each conductor are 
45 mm $\times$ 46 mm and 114, respectively.
The excitation curve of the coil was measured 
by a hall probe ( Group3 Technology, digital teslameter of DTM-151-DS, 
probe of MPT-141-10S, accuracy of $\pm $10$^{-4}$ ) and 
%compared to a TOSCA calculation in the Figure~\ref{fig:by_coil}. 
compared to a calculation by OPERA3D-TOSCA\cite{cite:tosca} which is 
a magnetic field calculation software by using 
three dimensional finite element method in the Figure~\ref{fig:by_coil}. 
%The measurement and TOSCA calculation are consistent, although the measured 
%correlation is slightly less than that of TOSCA calculation. This results are 
%reasonable. The measurement is done by the hall probe. If the probe has 
%small angles or/and little displacement from the center, displayed values should 
%be lower than the real center value of $B_{y}$.
The measured values of magnetic field of 
y-component ( B$_{y}$ ) are lower by a few $\%$ than those of 
TOSCA calculation. If the probe has                                                                      
small angles or/and little displacement from the center, 
the measured values could be lower than the real center value of $B_{y}$. 
In the present paper, the following values of magnetic field are derived 
from the applied current by using the measured relation.

\begin{figure}[htbp]
\begin{center}
\includegraphics[width=9cm]{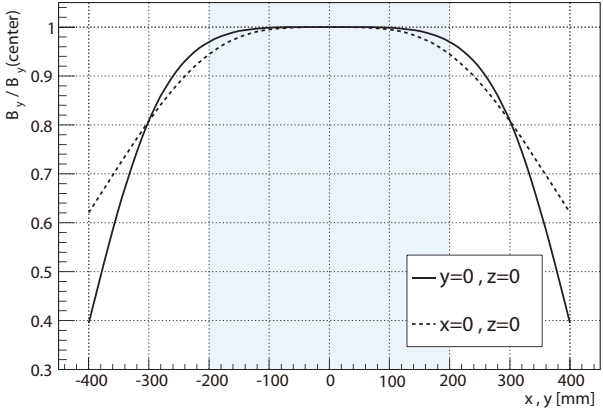}
\caption{$B_{y}/B_{y}$(center) vs. displacement from the center 
of the coils by the TOSCA calculation. The solid line describes the value when 
the values of $x$ and $y$ are fixed as 0 mm. The dashed line describes 
the value when the values of $x$ and $z$ are fixed as 0 mm. 
The calculation is consistent with the measurement within required accuracy for the bucking coil test.}
\label{fig:by_ratio}
\end{center}
\end{figure}

% The range of is an area?  Why not a volume?
The uniformity of the Helmholtz coil was also measured.
In the range of 400 mm $\times $ 400 mm around the center which 
we put PMT for the test, 
the uniformity of $B_{y}$ was 
\begin{equation}
\frac{B_{y}}{B_{y}\rm{(center)}} > 0.94
\label{eq:uniformity}
\end{equation}
The uniformity is checked by TOSCA as well.
Figure~\ref{fig:by_ratio} is showing $\frac{B_{y}}{B_{y}{\rm(center)}}$ by 
TOSCA calculation. The solid line describes the value when 
the values of $x$ and $y$ are fixed as 0 mm. The dashed line describes 
the value when the values of $x$ and $z$ are fixed as 0 mm. 
As seen in Figure~\ref{fig:by_ratio}, the TOSCA calculation 
shows $\frac{B_{y}}{B_{y}\rm{(center)}} > 0.94$ in the 
range of 400~mm $\times$ 400~mm, which is consistent with the measurement 
within required accuracy for the bucking coil test.
%The uniformity by the TOSCA calculation and that of measurement are 
%very consistent.

The other components of magnetic field in the region of 
400~mm $\times$ 400~mm were
\begin{equation}
\frac{B_{x,z}}{B_{y}\rm{(center)}} < 0.03
\end{equation}
according to the measurements.

\subsection{Light generation and circuit}
An LED (STANLEY 3889S Yellow) was used as the light source 
for the gain study of PMT. 
The peak wave length and luminous intensity of the LED 
are 690 nm and 4 mCd. 
The LED was covered by a light diffusion cap in order to achieve
uniform illumination of the PMT photocathode.
The LED was placed at the bottom of a paper cylinder with an
inside surface covered by a teflon reflector sheet.  The 
distance between the bottom of the paper cylinder and the 
photocathode of PMT was 260 mm. The cylinder was wrapped 
with two turns of light-tight black sheets to prevent light leaks.
\begin{center}
\begin{figure}[htbp]
\includegraphics[width=8cm]{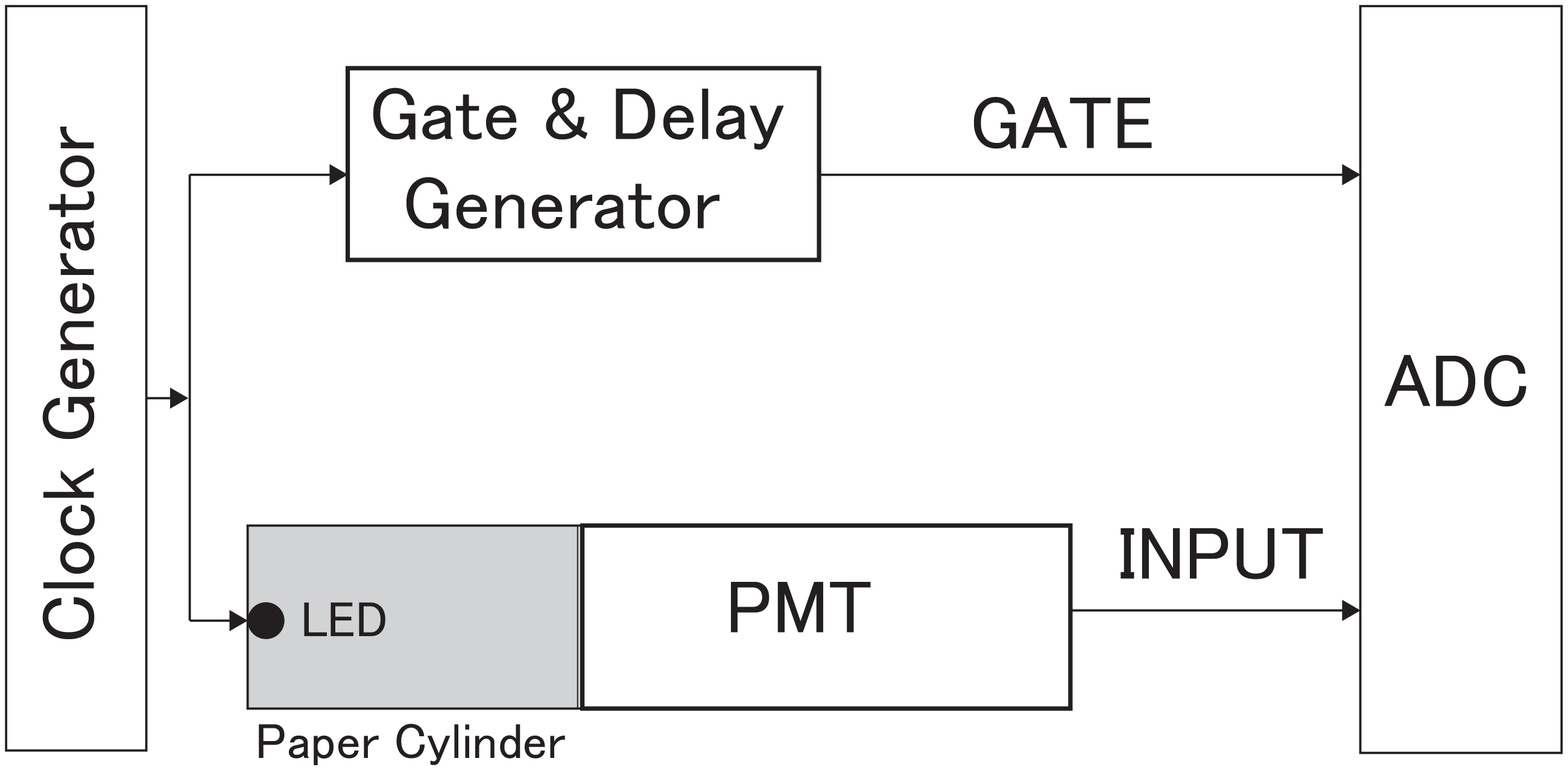}
\caption{Electronics block diagram for the bucking coil PMT tests.}
\label{fig:circuit}
\end{figure}
\end{center}

Figure~\ref{fig:circuit} shows the electronics block diagram for the PMT tests.
The LED generates light using Transistor-Transistor Logic ( TTL ) pulses from a Clock Generator.
The light intensity of the LED is controlled by varying the width of the pulses.
The width was set at $\sim $30~ns to generate $\sim $200 photo-electrons.
The charge information from the PMT was digitized by an 
Analog to Digital Converter ( ADC ) with a trigger
derived from the LED drive signal.

%%%%%%%%%%%%%%%%%%%%%%%%%%%%%%%%%%%%%%%%%%%%%%%%%%%%
%\section{Results}
\subsection{Effect of magnetic field on PMT}
Studies were made of the effects of magnetic fields on the PMTs
before the bucking coils were installed.
The results are presented as ``relative gain'', which is defined as the 
ratio of ADC value with a magnetic field 
to the ADC value with no magnetic field.  It should be noted that any
% change in ``relative gain'' can be a combination of changes in the number
change in ``relative gain'' would originate from multiple sources as 1) changes in the number
of photoelectrons collected by the first dynode of the PMT and 2) changes in
the amplification factor after the first dynode of the PMT.
%SAW. Perhaps relative gain should be changed to relative yield

%%%%%%%%%%%%%%%%%%%%%%%%%%%%%%%%%%%%%%%%%%%%%%%%%%%%
\subsubsection{Angular dependence}
The $\theta$ is defined as the angle between the magnetic field
direction ( the $y$-axis ) and the PMT 
axis ( perpendicular to the face of the PMT ) of the PMT.
\begin{figure}[!htbp]
\includegraphics[width=9cm]{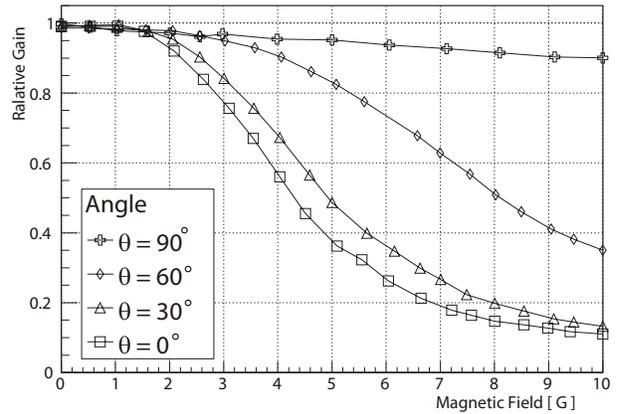}
\caption{Relative gain of PMT as a function of magnetic field.
PMT HV is set to $-$2000~V.  Different markers represent 
different angles between the tube axis and magnetic field direction ( $\theta = 0 , 30 , 60 , 90 $ degrees ).}
\label{fig:PMT_ang_dep}
\end{figure}
Figure~\ref{fig:PMT_ang_dep} shows the 
angular dependence of the relative gain. 
The relative gain is more than $90\%$ with $B_{y} < 2$ G for 
all $\theta$. However, it
drops to $\sim $40$\%$ with a magnetic field of 5~G parallel 
to the PMT axis ( $\theta = 0^{\circ}$ ). 
5~G is the magnitude of the fringe
field seen on these tubes in the spectrometer. 
As seen, the relative gain is strongly affected by magnetic fields along
the axis of the PMT.
%%%%%%%%%%%%%%%%%%%%%%%%%%%%%%%%%%%%%%%%%%%%%%%%%%%%
\subsubsection{Effect on one photo-electron detection}
\label{sec:ope}
\begin{figure}[htbp]
\includegraphics[width=9cm]{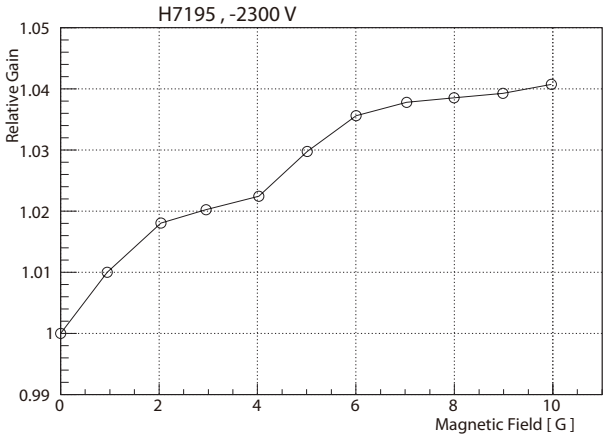}
\caption{The relative gain of one photo-electron as a function of magnetic field. 
The applied HV on PMT ( H7195 ) is $-$2300~V .}
\label{fig:PMT_opemag_dep}
\end{figure}

The relative gain when using the LED is strongly affected by the 
magnetic field as discussed in the previous section. 
To see changes in the relative gain for a single photo-electron peak, 
% the trigger was changed to a self trigger.
data were taken with a self trigger.
Figure~\ref{fig:PMT_opemag_dep} shows relative gain of 
one photo-electron peak as a function of magnetic field.
The relative gain of one photo-electron is near to 1, and is slightly 
increasing with field.
If the overall relative gain reduction originates only from 
that gain factor reduction after the first dynode, this result
should show similar behavior to figure \ref{fig:PMT_ang_dep} when we used an LED trigger.
%It means that no big effect on it from the magnetic field.
%In addition, the rate of the signal is reduced.  % What?  Out of place sentence.
In addition, a signal rate reduction was observed.  % What?  Out of place sentence.
According to these results, the 
reduction in relative gain originates mainly from a reduction of 
probability that electron which is converted from photon 
on photocathode reaches the first dynode of PMT.

It is also noted that the relative gain increase in 
figure~\ref{fig:PMT_opemag_dep} does not necessarily imply that 
gain factor is increased.  Under higher magnetic fields, only higher energy
electrons may be able to reach the first dynode.  These higher energy electrons
can knock off more secondary electrons at the first dynode, and 
what we observed might be biased by these events.
% Actually, if it is as I described it, this is an increase in gain
% at the expense of efficiency.

%%%%%%%%%%%%%%%%%%%%%%%%%%%%%%%%%%%%%%%%%%%%%%%%%%%%
\subsubsection{Applied HV dependence}
\begin{figure}[htbp]
\includegraphics[width=9cm]{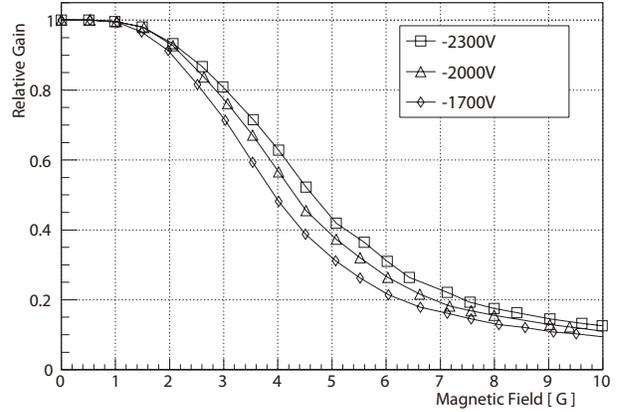}
\caption{The relative gain as a function of the magnetic field 
with differnt HV settings of $-$2300~V , $-$2000~V and $-$1700~V for the PMT ( H7195 ). }
\label{fig:PMT_HV_dep}
\end{figure}
Figure~\ref{fig:PMT_HV_dep} shows the relative gain 
as a function of the magnetic field which is applied
parallel to the axis of the PMT ( $\theta = 0^{\circ}$ ) 
with different HV settings. As the high voltage 
is increased, the effects of the magnetic 
field are eased slightly (e.g. the relative gain at 
5~G are 0.42 for $-$2300~V and 0.31 for $-$1700~V). 
However, this improvement is insufficient for 
our experimental requirements.

%%%%%%%%%%%%%%%%%%%%%%%%%%%%%%%%%%%%%%%%%%%%%%%%%%%%
%\subsubsection{mini summary}
%%%%%%%%%%%%%%%%%%%%%%%%%%%%%%%%%%%%%%%%%%%%%%%%%%%%
\subsection{Bucking coil test}
\begin{figure}[!htbp]
\begin{center}
\includegraphics[width=7cm]{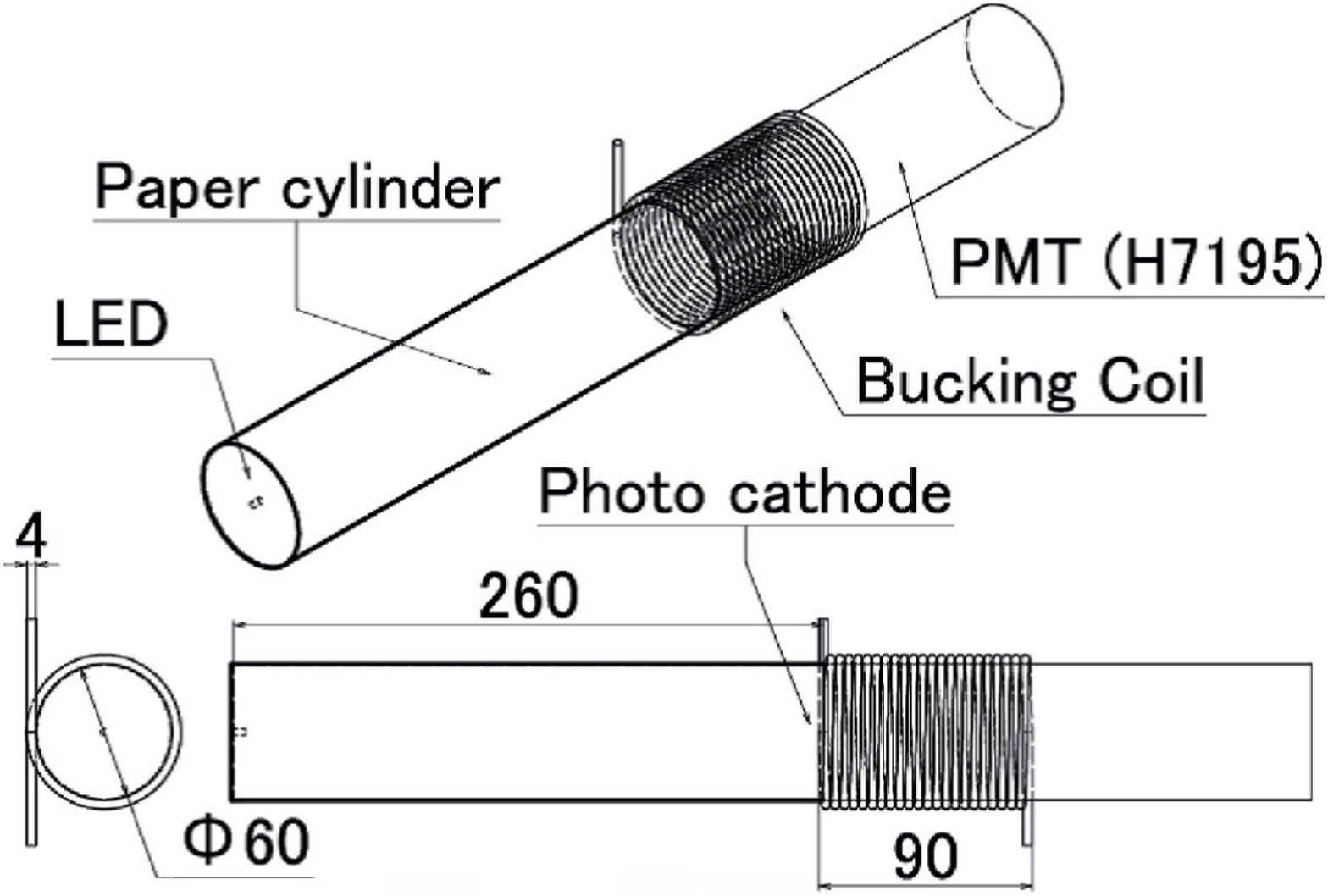}
\caption{The setup of the bucking coil on PMT for the test. 
The bucking coil of 20 turns was wound around the PMT, starting at the
photocathode and extending for 90~mm.  Unit in the figure is mm.}
% Why not redo the figure with cm instead of mm to be consistent?
\label{fig:buckingcoil_setup}
\end{center}
\end{figure}
Figure~\ref{fig:buckingcoil_setup} shows 
setup of a bucking coil on a PMT as used for the tests.
A wire with cross section of 4~mm$^{2}$ and
a resistance of 0.2~$\Omega /$m was used to wind the bucking coil.
The wire of 20 turns was rolled onto the PMT, starting at photocathode making a
coil with the length of 90~mm.
The current applied to the bucking coil were set to cancel the 
magnetic field inside the PMT.

\begin{figure}[!htbp]
\includegraphics[width=9cm]{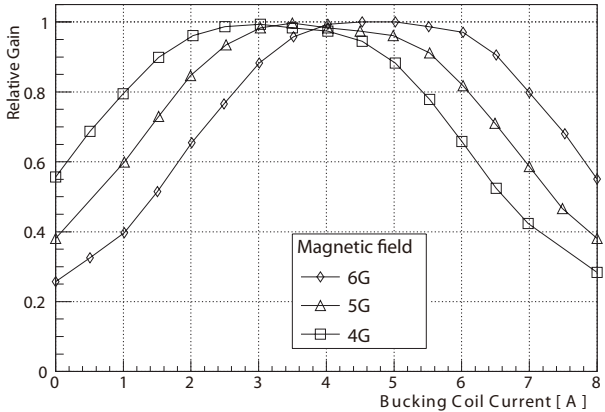}
\caption{Relative gain as a function of the bucking coil 
current under magnetic fields of 4~G, 5~G and 6~G ( $\theta = 0^{\circ}$ ) 
%The bucking coil length was 90~mm starting at the photocathode of 
The length of bucking coil of 20 turns was 90~mm starting at the photocathode of 
PMT ( H7195 , $-$2000~V ).}
\label{fig:BC}
\end{figure}

The relative gain as a function of current for the bucking coil under 
magnetic fields of 4~G, 5~G and 6~G ( $\theta = 0^{\circ}$ ) 
are shown in the Fig.\ref{fig:BC}.
The relative gains were recovered to $\sim$1 for 
each applied magnetic field on the PMT by choosing the optimal 
current for the bucking coil. Those curves have a plateau 
region. If one chooses bucking coil current of 2.5~A $\sim $ 5.0~A, 
the relative gain is more than 0.95 under the magnetic field of 5~G.
%for the lower and higher fields used.

%%%%%%%%%%%%%%%%%%%%%%%%%%%%%%%%%%%%%%%%%%%%%%%%%%%%
\subsubsection{Bucking coil position dependence}
\begin{figure}[!htbp]
\includegraphics[width=9cm]{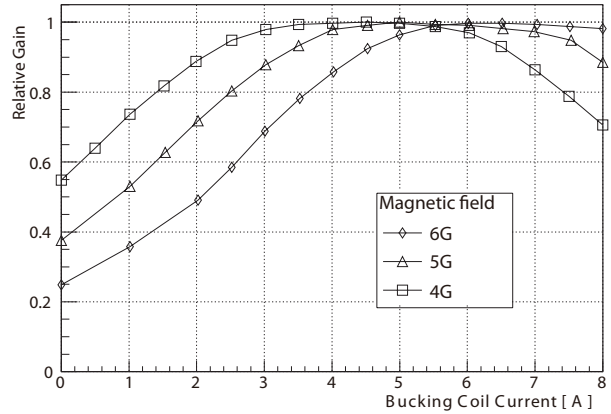}
\caption{Relative gain as a function of the 
bucking coil current under magnetic fields of 4~G, 5~G and 6~G.
The bucking coil started 40~mm away from the photocathode of the
PMT ( H7195 , $-$2000~V ).}
\label{fig:BC_pos_dep}
\end{figure}
In the experiment, JLab E05-115, bucking coil cannot be rolled 
onto the PMT starting at the photocathode because of physical interferences
from the frame of the {\cerenkov} detector. 
In anticipation of this constraint, tests were made with
a bucking coil that started 40~mm away from the photocathode.
The results are represented in the Fig.\ref{fig:BC_pos_dep}. 
The relative gains are recovered to $\sim$~1 for each 
applied magnetic field as well as the Fig.~\ref{fig:BC}.
% What does ``as well as the Fig.~\ref{fig:BC} mean? 
%However, the currents for the bucking coil which 
%are needed to recover the relative gain
%up to $\sim$~1 are higher compared to those of Fig.~\ref{fig:BC}.
However, the necessary currents are higher 
than those of Fig.~\ref{fig:BC}.
%This means that magnetic field around the photocathode of the PMT 
%affects more on the relative gain.
This means that the relative gain is more sensitive to 
the magnetic field around the photocathode of the PMT.

%%%%%%%%%%%%%%%%%%%%%%%%%%%%%%%%%%%%%%%%%%%%%%%%%%%%
\subsubsection{Angular dependence}

\begin{figure}[!htbp]
\includegraphics[width=9cm]{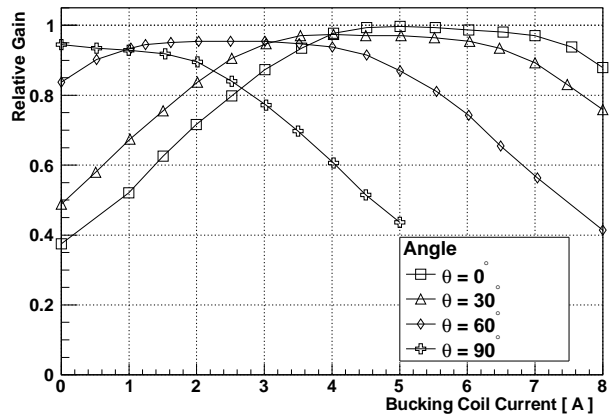}
\caption{Relative gain as a function of the bucking coil current under 
a magnetic field of 5~G for the angles of $\theta = $0, 30, 60 and 90 degrees 
( H7195 , $-$2000~V ).}
\label{fig:BC_ang_dep}
\end{figure}

Figure~\ref{fig:BC_ang_dep} shows the relative gain as 
a function of the bucking coil current with
angles of $\theta =$ 0, 30, 60 and 90 degrees. 
As angles of the magnetic field with respect to 
%the axis of PMT closer to 90 degrees, 
the axis of PMT closer to $90^{\circ}$, 
the maximum relative gains are lower though they are still 
more than 0.95 for $\theta = 90^{\circ}$.   % What?  Makes no sense.
%This is because some component of magnetic field which 
This imperfection of relative gain recovery is considered 
to be caused by a perpendicular component of magnetic field 
to the axis of PMT which cannot be cancelled by the 
bucking coil.
%cannot be cancelled by the bucking coil still
%remains.  % at x degrees?
%%%%%%%%%%%%%%%%%%%%%%%%%%%%%%%%%%%%%%%%%%%%%%%%%%%%
\subsubsection{Number of turns dependence}
\begin{figure}[!htbp]
\includegraphics[width=9cm]{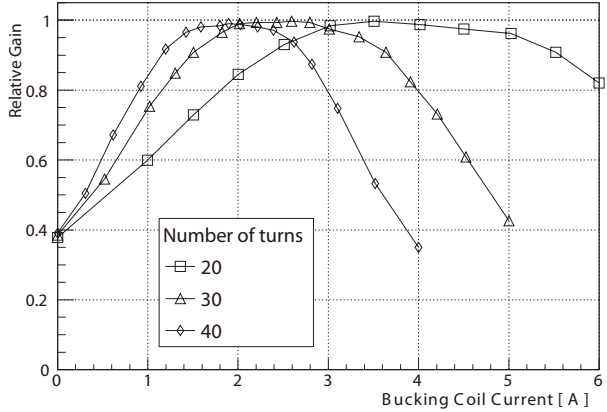}
\caption{Relative gain as a function of bucking coil current 
with different number of bucking coil turns. 
A magnetic field of 5~G ( $\theta = 0^{\circ}$ ) 
was used, and the bucking coil 
started at the photocathode of the PMT ( H7195, $-$2000~V ).}
\label{fig:BC_turns_dep}
\end{figure}
A relation between relative gain and the number of bucking coil turns 
is shown in Fig.~\ref{fig:BC_turns_dep}. 
In this test, a magnetic field of 5~G was applied on PMT ( $\theta = 0^{\circ}$ ).
The bucking coil started at the photocathode of the PMT, 
and a HV setting of $-$2000~V was used.
Less current is needed to recover the relative gain up to 
$\sim $1 with coils with a greater number of turns though 
the plateau region becomes narrower.
% Isn't this obvious?  Field roughly proportional to number of turns

%%%%%%%%%%%%%%%%%%%%%%%%%%%%%%%%%%%%%%%%%%%%%%%%%%%%
\subsubsection{HV dependence}
\begin{figure}[!htbp]
\includegraphics[width=9cm]{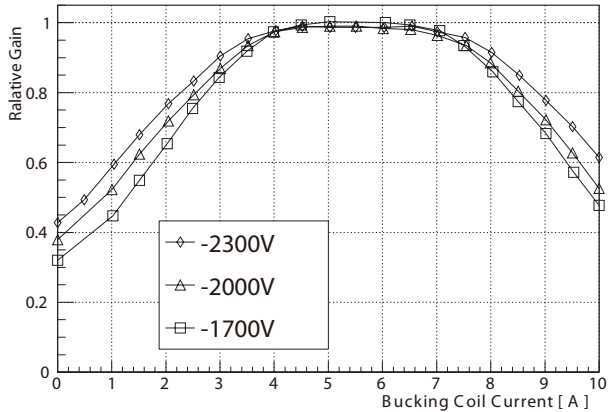}
\caption{Relative gain as a function of the bucking coil 
current with different HV settings, $-$1700~V, $-$2000~V and $-$2300V. 
A magnetic field of 5~G ( $\theta = 0^{\circ}$ ) 
was applied, and the bucking coil 
position started at the photocathode of the PMT ( H7195 ).}
\label{fig:BC_hv_dep}
\end{figure}
Figure~\ref{fig:BC_hv_dep} shows the relative gain as a function 
of bucking coil current with different HV settings, $-$1700~V, $-$2000~V and 
$-$2300~V. 
In this test, a magnetic field of 5~G was applied to the PMT ( $\theta = 0^{\circ} $ ), 
%and the bucking coil of 20 turns was located starting at the photocathode of the PMT. 
and the bucking coil of 20 turns was wound starting at the photocathode of the PMT. 
The necessary currents to recover relative gains 
up to $\sim $~1 for each HV setting are very similar. 
In the actual experiment setup, different HV settings were used
for PMTs. 
The obtained results imply that a single bucking coil current can work 
though different HV settings are used in the actual experimental setup.
%These results imply that criteria other than recovery
%of relative gain can be used to set high voltages in the actual detector.

%%%%%%%%%%%%%%%%%%%%%%%%%%%%%%%%%%%%%%%%%%%%%%%%%%%%
\section{Bucking coil implementation for the experiment JLab E05-115}
%%%%%
\subsection{Water {\cerenkov} Detectors}
\label{sec:water_real}
\begin{figure}[!htbp]
\begin{center}
\includegraphics[width=9cm]{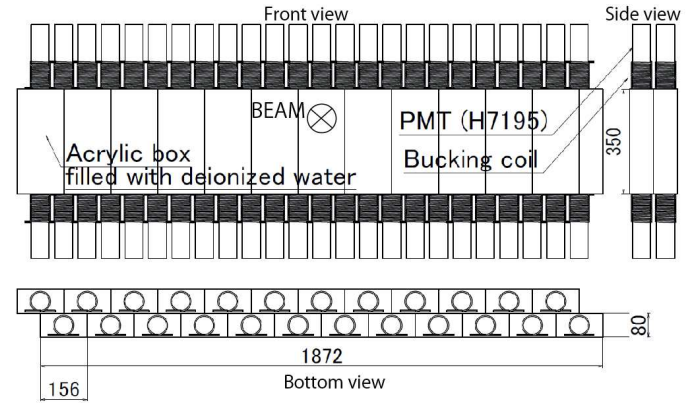}
\caption{A schematic drawing of the water {\cerenkov} detector. 
There are two layers with 12 segments for each layer. A bucking coil 
of 40 turns was rolled around each PMT, starting $\sim$40~mm away 
from the photocathode. Unit in the figure is mm.}
\label{fig:wc_setup}
\end{center}
\end{figure}
\begin{figure}[!htbp]
\begin{center}
\includegraphics[width=7cm]{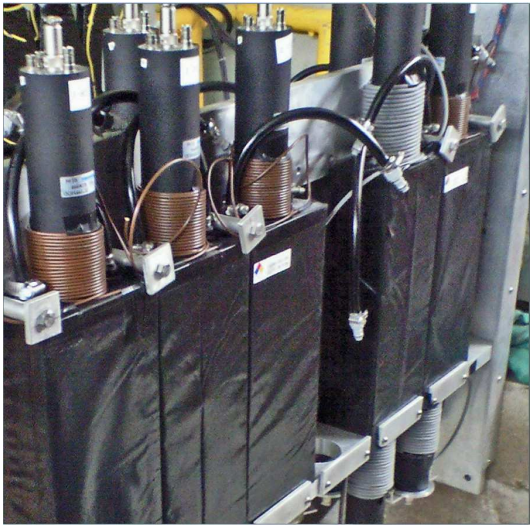}
\caption{A photograph of the bucking coils on the PMTs of water {\cerenkov} detectors. 
A bucking coil of 40 turns was used for each PMT.}
\label{fig:wc_bg_photo}
\end{center}
\end{figure}

A schematic drawing of the water {\cerenkov} 
detector is shown in the Fig.~\ref{fig:wc_setup}. 
There are two layers of water {\cerenkov} detector 
with 12 segments for each layer. 
A detector segment consists of an acrylic box with its
inside surface covered by a teflon sheet, and PMTs ( H7195 ) on 
top and bottom of the box with UV-glass windows. 
Deionized water ( resistivity of 18 M$\Omega \cdot$cm, 
%$n=1.33$ was used as the radiation medium. 
the refraction index of 1.33 ) was used as the radiation medium. 
A bucking coil of 40 turns was rolled around each PMT, starting $\sim$40~mm away from the 
photocathode.
Figure~\ref{fig:wc_bg_photo} is a photograph of bucking coils on PMTs 
of the water {\cerenkov} detectors. 

\begin{figure}[!htbp]
\includegraphics[width=9cm]{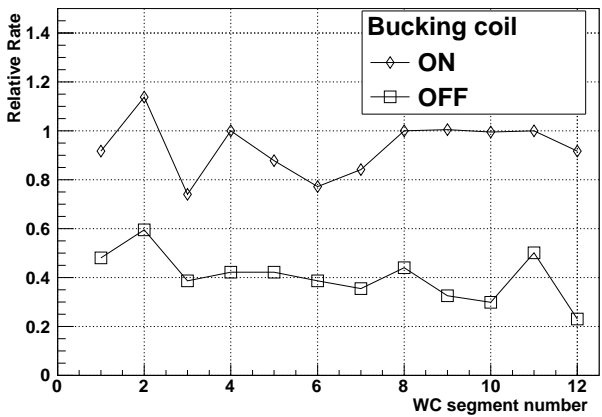}
\caption{Relative rates of each segment of the first layer of water {\cerenkov} detectors 
when the HKS dipole magnet is on. The relative rates are recovered up to 
$\sim$1 when the bucking coils are ON ( 40 turns , the current of 2~A ).}
\label{fig:wc_real}
\end{figure}

In the experiment, the effects of bucking coil 
for the water {\cerenkov} detector were checked by 
measuring the counting rates of each PMT, which is 
equivalent to measuring the relative gains as discussed in section \ref{sec:ope} 
if the discriminator threshold is kept the same. 
The ratios of counting rates of each PMT when the HKS dipole 
magnet was on to those when the magnet was off are shown in 
the Fig.~\ref{fig:wc_real}. 
%One of the sets of ratios are for
%when the current for bucking coil was OFF, while the other is for
%when a bucking coil current of $\sim$~2~A was applied. 
The counting rates for each PMT are decreased by more than $\sim$50$\%$ 
when the bucking coil is not used. On the other hand, the counting rates with 
bucking coil ON ( 40 turns, the current of 2~A ) are recovered to those with
no HKS magnetic field.

%%%%%
%\subsection{The effect of the Bucking coil}
\subsection{Aerogel {\cerenkov} detectors}
\label{sec:aerogel_real}
\begin{figure}[!htbp]
\begin{center}
\includegraphics[width=9cm]{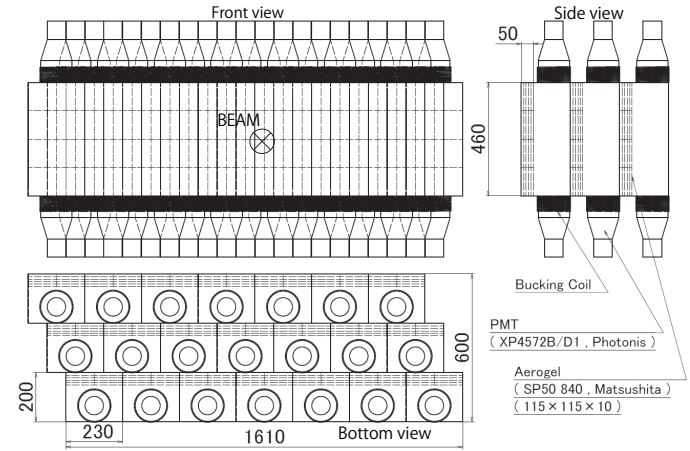}
\caption{A schematic drawing of the aerogel {\cerenkov} detector. There are 
three layers with 7 segments for each layer. Unit in the figure is mm.}
\label{fig:ac_setup}
\end{center}
\end{figure}

\begin{figure}[!htbp]
\begin{center}
\includegraphics[width=9cm]{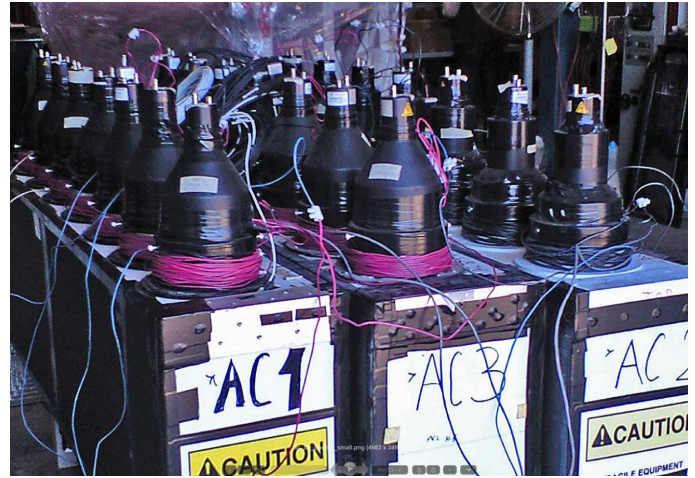}
\caption{A photograph of the aerogel {\cerenkov} detectors. Bucking coils 
are rolled around the photocathodes of each PMT.}
%Bucking coil on PMT for the aerogel {\cerenkov} detector.}
\label{fig:ac_bg_photo}
\end{center}
\end{figure}
Figure~\ref{fig:ac_setup} shows a schematic drawing of the 
aerogel {\cerenkov} detector. There are three layers of aerogel {\cerenkov} 
detectors, and each layer has 7 segments.  A segment consists of a 
structurally strengthened paper box, 
40 pieces of hydrophobic silica aerogel 
tiles ( SP-50 , Matsushita\footnote{Currently produced by the Japan Fine Ceramics Center} , 
refraction index of 1.05 ) and two PMTs 
( XP4572B/D1, Photonis, photocathode of bialkali, dynode stages of 10, 
typical supply voltage of +2100~V, typical gain of 2.0$\times$10$^{7}$ ), 
located on the top and bottom of the box without any windows. 
It is noted that the PMT which is used for aerogel detector is more 
sensitive to magnetic fields because its diameter ( 5 inches ) is larger than that for 
water {\cerenkov} detector ( 2 inches ). 
%\cortext[cor4]{Japan Fine Ceramics Center}

\begin{figure}[!htbp]
\includegraphics[width=9cm]{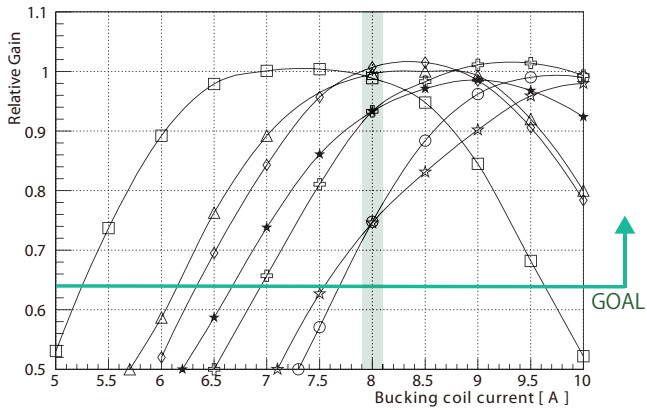}
\caption{Relative gain of each segment of the first layer of aerogel {\cerenkov} detectors. 
In the experiment, a single current of 8~A was chosen for all PMTs.}
\label{fig:ac_real}
\end{figure}

Figure~\ref{fig:ac_bg_photo} shows the bucking coils on PMTs for the aerogel {\cerenkov} 
detectors. 40 turns of bucking coils were used for each PMT. 
Figure~ \ref{fig:ac_real} shows the relative gain of the 
each PMT in the first layer of the aerogel {\cerenkov} detectors as a function of 
bucking coil current. 
%Although bucking coil currents to recover the relative gain up to 
%$\sim$~1 are different for each segment because leaked magnetic filed 
%of HKS dipole magnet is not uniform around the aerogel {\cerenkov} detector, 
%the ADC mean values for every PMT can be recovered by the bucking coil. 
%The bucking coil current of 8~A is selected to recover the relative gain for 
%each PMT as fairly as possible. 
Because the fringe field of the HKS dipole is not uniform in the region
of the aerogel {\cerenkov} detector, the optimal bucking coil currents
are different for each segment.  
%However, a single current of 8~A does a
%reasonable job of recovering most of the relative gain for all the PMTs.
However, a single current of 8~A was chosen since 
it recovered most of the relative gain reasonably for all PMTs.

%%%%%
%\subsection{$K^{+}$ Identfication after bucking coil implementation}
\subsection{$K^{+}$ identfication performance}
A schematic drawing of the HKS detectors is 
represented in the Fig.~\ref{fig:hks_setup}. 
The HKS consists of three layers of plastic scintillation detectors ( KTOF1X , KTOF1Y , KTOF2X ) 
for TOF ( Time Of Flight ) measurement, two layers of horizontal drift chambers ( KDC1 , KDC2 ) 
for particle tracking, and the aerogel and water {\cerenkov} detectors 
(AC1 , AC2 , AC3 , WC1 , WC2 ) for particle identification. 
%( AC , WC ) for particle identification. 
\subsubsection{On-line $K^{+}$ identification}
%The HKS trigger can be described by the following expression,
The HKS trigger was made by the following logical condition,
\begin{eqnarray}
{\rm HKS}_{ {\rm trigger} } =  {\rm CP}_{{\rm trigger}} \otimes K_{{\rm trigger}}
\label{eq:hks_trigger}
\end{eqnarray}
where, 
\begin{eqnarray}
{\rm CP}_{{\rm trigger}} = {\rm KTOF1X} \otimes {\rm KTOF1Y} \otimes {\rm KTOF2X},
\label{eq:cp_trigger}
%K_{{\rm trigger}} = \overline{ (AC1 \otimes AC2 \otimes AC3)}  \otimes (WC1 \otimes WC2)\\
%K_{{\rm trigger}} = \overline{ {\rm AC} }  \otimes {\rm WC}
%\label{eq:k_trigger}
\end{eqnarray}
\begin{eqnarray}
K_{{\rm trigger}} = \overline{ {\rm AC} }  \otimes {\rm WC}.
\label{eq:k_trigger}
\end{eqnarray}
CP$_{{\rm trigger}}$ in eq.(\ref{eq:cp_trigger}) 
is a charged particle trigger which consists of the coincidence of 
three layers of TOF detectors ( KTOF1X , KTOF1Y and KTOF2X ). 
In eq.~(\ref{eq:k_trigger}), AC represents 
%( Top PMT or Bottom PMT )$_{{\rm AC}1} \otimes$ ( Top PMT or Bottom PMT)$_{{\rm AC2}}$, 
(AC1 $\otimes$ AC2) $\oplus$ (AC2 $\otimes$ AC3) $\oplus$ (AC3 $\otimes$ AC1),  
and WC represent ( WC1 $\otimes$ WC2 ). The overline on AC means that 
the AC was used as veto to suppress $\pi ^{+}$. 
%The typical HKS trigger rate is $\sim$5~kHz with beam current of 
%$\sim$36~$\mu $A on $^{12}$C target ( 112~mg/cm$^{2}$ ). 
The typical HKS trigger rate was $\sim$1~kHz with a beam current of 
2~$\mu $A on a polyethylene target ( CH$_{2}$ , the material thickness of 450~mg/cm$^{2}$ ). 

Data with the HKS$_{{\rm trigger}}$ and the CP$_{{\rm trigger}}$ conditions were taken 
during the experiment although a coincidence between the HKS trigger ( HKS$_{{\rm trigger}}$ )
and the HES trigger was used for physics data. $\pi^{+}$ and $p$ rejection ratios  
were estimated by comparing the number of those events in the HKS$_{{\rm trigger}}$ 
and the CP$_{{\rm trigger}}$. The rejection ratios of $\pi^{+}$ and $p$
with a beam current of 2~$\mu $A on the polyethylene target 
%( CH$_{2}$ , $\sim$~450~mg/cm$^{2}$ ) 
were $7.4 \times 10^{-3}$ and $3.8 \times 10 ^{-2}$, respectively. 
%The $K^{+}$ survival ratio is 91.3 $\pm$ 4.5 $\%$ at this time.
The $K^{+}$ detection efficiency was 
%91.3 $\pm$ 4.5 $\%$.
91$\%$.
% Survival ratio implies surviving decay.
% What does ``at this time'' mean?

\subsubsection{Off-line $K^{+}$ identification}
\begin{figure}[!htbp]
\includegraphics[width=9cm]{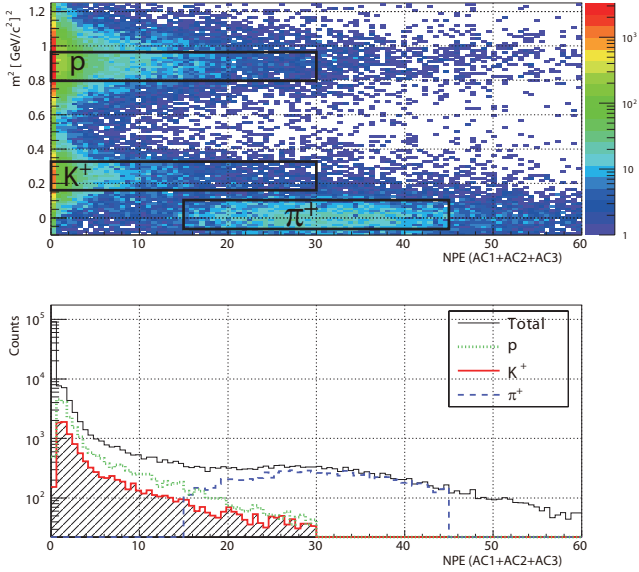}
\caption{Correlation between 
the number of photoelectrons summed over the three layers of the aerogel {\cerenkov} detector and 
mass squared (top), and x-projection of the top figure (bottom).}
\label{fig:msq_aer}
\end{figure}

At the on-line ( hardware ) trigger level, thresholds of the {\cerenkov} detectors
were set slightly loose so as to avoid over-cutting of 
$K^{+}$s. Therefore, some $\pi ^{+}$s and protons remained in the recorded data. 
The top plot of Fig.~\ref{fig:msq_aer} shows the correlation between 
number of photoelectrons from the sum of the three layers of 
aerogel {\cerenkov} detectors and mass squared of the measured particle. 
The bottom plot is an x-projection of the top graph. 
Mass squared, $m^{2}$ is calculated by the following equation,
\begin{eqnarray}
%m^{2} = \frac{p_{{\rm recon}}^{2}}{(1/\beta ^{2}) - 1}
m^{2} = \frac{k^{2}}{(1/\beta ^{2}) - 1}
\end{eqnarray}
where $k$ is the momentum reconstructed by transfer matrix of the 
HKS, and $\beta$ is the velocity factor derived from the TOF measurement. 
In Fig.~\ref{fig:msq_aer}, clusters of $\pi^{+}$, $K^{+}$ and $p$ 
are clearly seen. $\pi^{+}$s are rejected by applying the cut on
the number of photoelectrons in the aerogel {\cerenkov} detector. 

Two types of boxes of water {\cerenkov} detectors were used in the experiment. 
The main differences between them were reflective materials and PMT choice. 
On the low momentum side ( segment numbers from 1 to 6, BOX1 ), 
%the reflective material is white acrylic and PMT is 
white acrylic as reflective material and H7195 PMTs were used. 
%It was used in the JLab E01-011, too.
On the high momentum side where severer particle 
identification is required ( segment numbers from 7 to 12 , BOX2 ), 
%teflon sheets are the reflective material and the H7195UV tubes 
%which have higher efficiency to ultra violet light 
%than normal H7195 are used. 
teflon sheets as the reflective material and H7195UV PMTs
which has UV-glass window were used. H7195UV has the same 
responses to the magnetic field as H7195, but 
higher efficiency to ultra violet light.

\begin{figure}[!htbp]
\includegraphics[width=9cm]{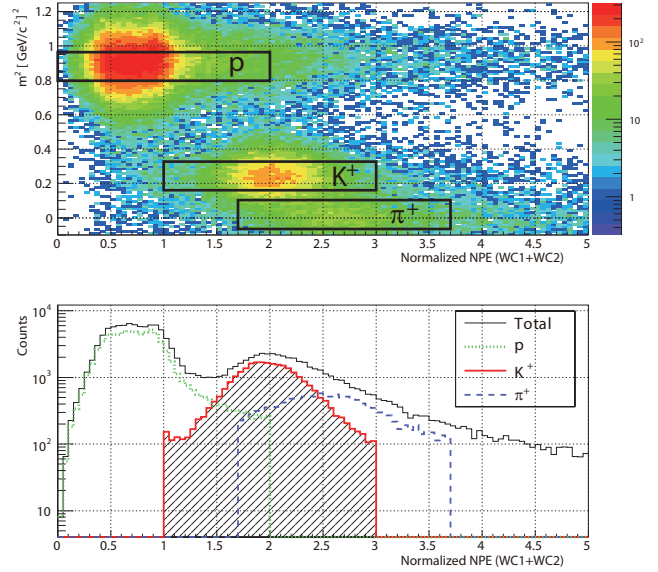}
\caption{Correlation between 
normalized number of photoelectrons from the sum of two layers of water {\cerenkov} detectors and 
mass squared (top), and the x-projection of the top figure (bottom). }
\label{fig:msq_wat}
\end{figure}

%BOX2 is newly developed for JLab E05-115. 
In cosmic ray tests, the average number of photoelectrons in BOX1 and
%BOX2 type segments were $\sim$40 and $\sim$100, respectively. 
BOX2 type segments were $\sim$50 and $\sim$100, respectively. 
There was difference of number of photoelectrons between 
BOX1 and BOX2, % You don't have to state that 40 is not equal to 100
and thus, normalized number of photoelectrons 
were introduced to adjust $K^{+}$ peak to 1.  % Which one was normalized?
The top plot of Fig.~\ref{fig:msq_wat} shows the correlation between 
the normalized number of photoelectrons from the sum of two layers of water {\cerenkov} detectors and 
the mass squared, while the bottom plot is the x-projection of the top figure. 
Protons and other particles ( $\pi^{+}$, $K^{+}$ ) can be 
separated by the cut on the number of photoelectrons. 

\begin{figure}[!htbp]
\begin{center}
\includegraphics[width=9cm]{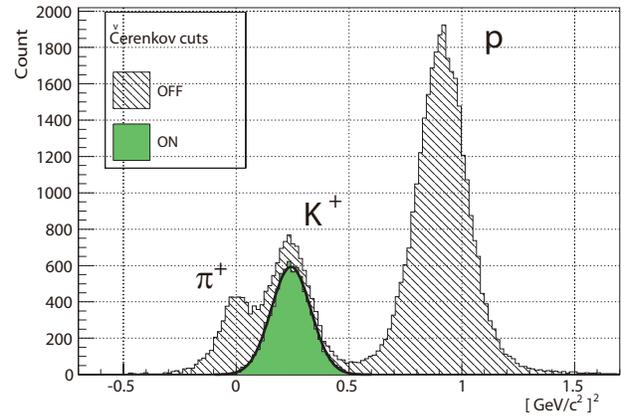}
\caption{Mass squared distribution before and after the cuts on the
{\cerenkov} detectors. $K^{+}$ events are clearly selected after the cuts. 
The width of single gaussian fitting for kaon peak is $\sigma \sim$~(~0.29~[GeV/$c^{2}$]~)~$^{2}$.}
\label{fig:msq}
\end{center}
\end{figure}

Figure \ref{fig:msq} shows the mass squared distribution before 
and after the cuts on the number of photoelectrons in the {\cerenkov} detectors. 
$K^{+}$ events are clearly selected after applying these cuts.
When the {\cerenkov} and mass squared cuts
were selected to keep 90$\%$ kaons in the total off-line events, 
%$<$2$\%$ and $<$1$\%$ events are misidentified as $p$ and $\pi^{+}$, respectively in 
%$<$2$\%$ and $<$1$\%$ events are  $p$ and $\pi^{+}$, respectively in 
%the off-line $K^{+}$ events ( CH$_{2}$ , $\sim$~450~mg/cm$^{2}$ ). 
%only 4$\%$ and 3$\%$ of the remaining events are $\pi^{+}$ and $p$, respectively
the contaminated $\pi^{+}$ and $p$ events were 4$\%$ and 3$\%$ 
of remaining events ( using the 450~mg/cm$^{2}$ CH$_{2}$ target ). 
In this case, the total ( on-line and off-line )
rejection powers of $\pi^{+}$ and $p$ 
were 6.5$\times$10$^{-4}$ and 6.1$\times$10$^{-5}$, respectively. 

% for  
%the offline remained $K^{+}$ events  (

\section{Summary}
The water and aerogel {\cerenkov} detectors, which were located 
a few meters away from the HKS dipole magnet, needed to be operated in
a fringe field of $\sim$5~G. 
The separation efficiency of $K^{+}$ from the other 
background particles ( $\pi^{+}$ and $p$ ) is deteriorated in these fields 
as the fields reduce the relative gains.
Monte Carlo simulations indicated that
relative PMT gains of more than 65$\%$ were required to
keep the yield of $\Lambda$ hypernuclei and the signal to noise ratio
reasonably high.
To operate the PMTs attached to {\cerenkov} radiators under this field,
bucking coils were installed to actively and locally
cancel the magnetic field on the PMTs.\\
~Tests with a Helmholtz coil 
were performed to study the effects of magnetic fields on the H7195 PMT
which was used for the water {\cerenkov} detectors. 
With a magnetic field of $\sim$5~G parallel to the axis of the PMT,
the gain was reduced by 60$\%$ ( relative gain of 0.40 ). 
A bucking coil of 40 turns and $\sim$2~A 
starting at the photocathode of PMT was 
needed to recover the relative gain up to $\sim$1 in this case. \\
~In JLab E05-115 experiment, 
bucking coils of 40 turns and current of 2~A were used for 
the water {\cerenkov} detectors, and coils with 40 turns and current of 
8~A were used for aerogel {\cerenkov} detectors. 
With these coils, a relative gain of more than 0.70 
and 0.75 for each segment of the water and aerogel {\cerenkov} 
detectors, respectively, were achieved.
The on-line rejection ratios of $\pi^{+}$ and $p$ with a beam current of 
2~$\mu$A on the CH$_{2}$ target ( the material thickness of 450~mg/cm$^{2}$ ) were
7.4 $\times 10^{-3}$ and 3.8 $ \times 10 ^{-2}$, respectively.
The efficiency for $K^{+}$ events was 
%91.3~$\pm$~4.5$\%$.  % at this time. 
91$\%$.  % at this time. 
Off-line, $K^{+}$ events were selected by applying cuts 
on the number of photoelectrons in the {\cerenkov} detectors. 
When {\cerenkov} and mass squared cuts were selected
to keep 90$\%$ efficiency in off-line kaon selection, 
%$<$2$\%$ and $<$1$\%$ events are misidentified as $p$ and $\pi^{+}$, respectively in 
%$<$2$\%$ and $<$1$\%$ of the remaining events were $p$ and $\pi^{+}$, respectively.
%the offline remained $K^{+}$ events  ( CH$_{2}$ , $\sim$~450~mg/cm$^{2}$ ). 
%only $\sim$4$\%$ and $\sim$3$\%$ of the remaining events were $\pi^{+}$ and $p$, respectively. 
the contaminated $\pi^{+}$ and $p$ events were 4$\%$ and 3$\%$ 
of remaining events ( using the 450~mg/cm$^{2}$ CH$_{2}$ target ). 
In this case, the total ( on-line and off-line ) rejection powers 
of $\pi^{+}$ and $p$ are 6.5$\times$10$^{-4}$ and 6.1$\times$10$^{-5}$, respectively. 
The implementation of bucking coils allowed for a very clean on-line and
off-line selection of $K^{+}$ events. 

\section*{Acknowledgements}
%We would like to thank the staff of JLab for supports for 
We would like to thank JLab staff of physics, accelerator and 
engineer divisions for supports for the experiments.
The program at JLab's Hall-C is supported by the Specially promoted
program (12002001), the Creative research program (16GS0201), 
%Scientific Research on Priority Areas (08239120) and Basic research (09304028, 09554007, 11440070, 15204014) of 
Grant-in-Aid by MEXT (15684005), Japan,
JSPS Research Fellowships for Young Scientists (244123), 
US-Japan collaboration research program, the JSPS Core-to-Core 
Program (21002) and the Strategic Young Researcher Overseas Visits Program for Accelerating 
Brain Circulation (R2201) by the Japan Society for Promotion of Science.
This work is supported by U.S. Department of Energy contract
DE-AC05-06OR23177 under which Jefferson Science Associates, LLC, operates
the Thomas Jefferson National Accelerator Facility.

%%%%%%%%%%%%%%%%%%%%%%%%%%%%%%%%%%%%%%%%%%%%%%%%%%%%

%\end{linenumbers}

%\appendix{
%The program at JLab's Hall C is supported by the Specially promoted program 
%(12002001), the Creative research program (16GS0201), Scientific Research on 
%Priority Areas (08239120) and Basic research (09304028, 09554007, 
%11440070, 15204014) of Grant-in-Aid by MEXT, Japan, US-Japan collaboration 
%research program, the JSPS Core-to-Core Program (21002) and 
%the Strategic Young Researcher Overseas Visits Program for Accelerating Brain 
%Circulation (R2201) by the Japan Society for Promotion of Science. 
%JLab is operated by The Southeastern Universities Research Association, 
%Inc. under U.S. DOE Contract No.DE-AC05-84ER40150.
%}



%% References



\begin{thebibliography}{00}

\bibitem{cite:proposal}O.Hashimoto, S.N.Nakamura, L.Tang, J.Reinhold et al., JLab E05-115 proposal (2005)
\bibitem{cite:osamu}O.Hashimoto et al., Nuclear Physics A835 , 121-128 (2010)
%\bibitem{cite:toshi}T.Gogami et. al. , Few-Body Systems ,  10.1007/s00601-012-0397-z (2012)
\bibitem{cite:lulin}L.Yuan et al., Physical Review C73 , 044607 (2006)
\bibitem{cite:tang}L.Tang et al., Nuclear Physics A790 , 679c-682c (2007)
\bibitem{cite:osamu2}O.Hashimoto et al., Nuclear Physics A804 , 125-138 (2008)
%\bibitem{cite:nue}S.N.Nakamura et al. , arXiv:1207.0571 (2012)
\bibitem{cite:nue}S.N.Nakamura et al., Physical Review Letters 100 , 012502 (2013)
%\bibitem{cite:tosca}http://www.cobham.com/about-cobham/aerospace-and-security/about-us/antenna-systems/specialist-technical-services-and-software/products-and-services/design-simulation-software/opera/opera-3d.aspx
\bibitem{cite:tosca}http://www.cobham.com , Cobham Antenna Systems, Vector Fields Simulation Software
\end{thebibliography}



\end{document}